\documentclass[english,aps,preprintnumbers]{revtex4}

\usepackage{babel}
\usepackage{amsmath,amssymb}
\usepackage{graphicx}

\newcommand{\lnd}{\ln 2\delta}
\newcommand{\lndd}{\ln^2 2\delta}
\newcommand{\del}[1]{\,\delta^{#1}}
\newcommand{\order}[1]{\mathcal{O}\left( #1 \right)}

\begin{document}

\preprint{Alberta Thy 11-08}

\title{Semileptonic decays in the limit of a heavy daughter quark}

\author{Matthew Dowling, Jan H. Piclum, and Andrzej Czarnecki}

\affiliation{Department of Physics, University of Alberta, Edmonton, Alberta,
Canada T6G 2G7}

\begin{abstract}
The rate of the semileptonic decay $b\to c\ell\nu$ is calculated with
$\order{\alpha_{s}^{2}}$ accuracy, as an expansion around the limit
of equal masses of the $b$ and $c$ quarks. Recent results obtained
around the limit of the $c$-quark much lighter than $b$ are confirmed.
Details of the new expansion method are described.
\end{abstract}
\maketitle

\section{Introduction}
Very recently, next-to-next-to-leading order (NNLO) QCD corrections to
the semileptonic $b\to c$ decay were calculated with full account of
the charm quark mass \cite{Melnikov:2008qs,Pak:2008qt} (see also
Ref.~\cite{Pak:2008cp}). The former paper employed a numerical method
while in the latter the decay rate was expanded in the ratio of quark
masses, $\rho=m_{c}/m_{b}$. These two approaches are complementary,
with the numerical one having better accuracy for the larger daughter
quark mass and the analytical expansion being obviously better for a
lighter one.  The physically most interesting is the region of the
actual quark mass ratio, $\rho\sim0.25..0.3$.  Both methods are
applicable in this domain and agree very well with each other. The
resulting prediction for the $b$-quark decay rate will improve the
accuracy of the quark mixing parameter $V_{cb}$.

In the present paper we provide an additional check of that QCD
correction.  We construct an analytical expansion like in
\cite{Pak:2008qt}, but around the opposite limit: instead of starting
with $\rho=0$ (massless charm) we expand around $\rho=1$ (equally
heavy $b$ and $c$ quarks).  We find that this leads to a faster
convergent series whose sum smoothly matches that found in
\cite{Pak:2008qt}. As a result we now have a set of analytical
expressions valid in the whole range of possible quark masses. In
addition, the method of asymptotic expansions is applied to a new
kinematic configuration.

\section{Calculational Method for the Decay Rate}
\begin{figure}
  \includegraphics[width=\textwidth]{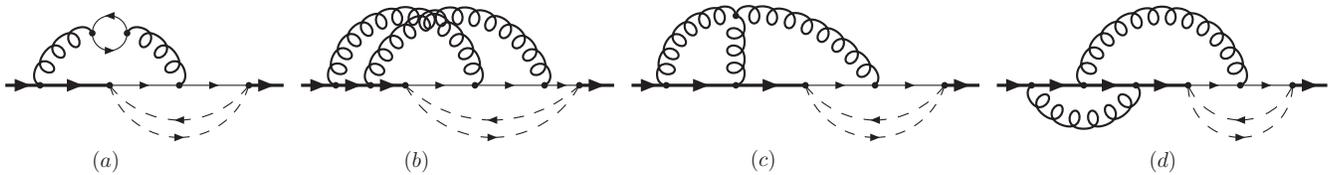}
  \caption{\label{fig::dias}Sample diagrams contributing to the decay
    width. Thick and thin lines denote $b$ and $c$ quarks, respectively. Curly
    lines denote gluons and dashed lines denote leptons. All quark flavors
    have to be considered in the closed loop.}
\end{figure}

Using the optical theorem, we calculate the decay width from the
imaginary parts of $b$-quark self-energy diagrams up to four loops,
such as shown in Fig.~\ref{fig::dias}.  These diagrams contain two
masses, $m_b$ and $m_c$, and it is not known how to compute them
analytically.  We thus treat the mass difference $m_b-m_c=m_b(1-\rho)$
as a small quantity and construct an expansion around the limit of
equal masses.  The expansion parameter is $\delta \equiv 1 - \rho$.

This expansion is peculiar in the sense that the decay is not
possible at the limiting point, $\delta=0$. This leads to a
strong suppression of our result as $\rho$ tends to one and, as will
be seen below, ensures good convergence of the expansion. Furthermore,
there are no contributions from the region where all loop momenta are of order
$m_b$. This makes the calculation significantly simpler than the
complementary expansion around $\rho=0$ \cite{Pak:2008qt}.

A somewhat similar configuration was considered in
\cite{Czarnecki:2001cz}, where the decay $b\to u\ell \nu$ was
evaluated near the limit of the maximum invariant mass of the
leptons.  The difference in the present case is that it is the
daughter quark that is massive and almost saturates the phase space.
Since that massive quark radiates, the calculation is more involved.
We explain it in some detail below.

\begin{figure}
  \centering
  \includegraphics[width=0.6\textwidth]{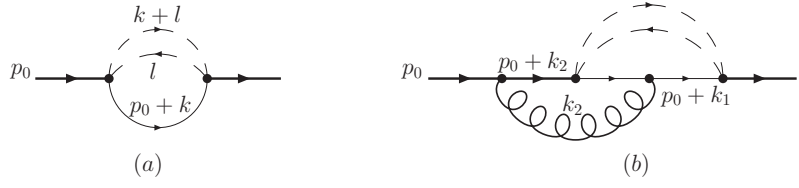}
  \caption{\label{fig:tree} In $(a)$ the tree level diagram for the decay
    $b\to c\ell \nu$ is depicted, while $(b)$ shows the general topology of
    the first order correction. Thin and thick lines indicate $c$ and  $b$
    quarks, respectively. Dashed lines denote leptons and curly lines denote
    gluons.}
\end{figure}

In the first step we integrate over the loop momentum in the massless
neutrino-charged lepton loop, replacing it with a fractional power of
the momentum flowing through it, $1/k^{2\epsilon}$ where
$D=4-2\epsilon$ is the number of dimensions in dimensional
regularization.  For example, Fig.~\ref{fig:tree} $(a)$ shows the
lowest-order diagram; the lepton-loop momentum is $l$.  

\begin{figure}
\begin{tabular}{lll}
\parbox[c][1\totalheight]{3cm}{%
Full (unexpanded) diagram%

\includegraphics[width=2.9cm]{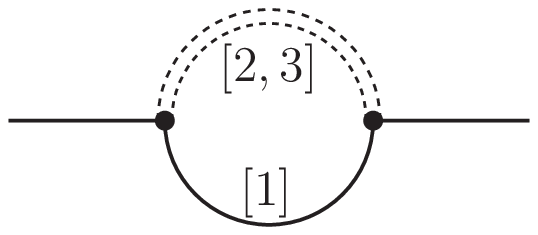}} & \multicolumn{2}{l}{%
\parbox[c][1\totalheight]{5cm}{%
$[1] = (p_0 + k)^{2}+m_c^{2}$ \, $[2,3] = k^{2\epsilon}$
}}\\[8ex]
\parbox[c][1\totalheight]{3cm}{%
 Region 1 %

\includegraphics[width=2.9cm]{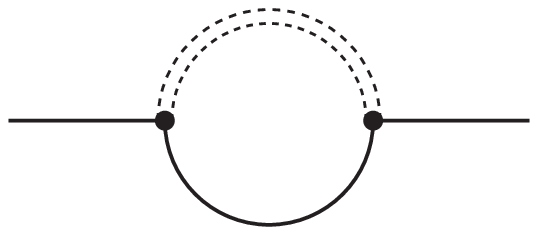}} & %
\parbox[c][1\totalheight]{2cm}{%
$|k| \sim m_b$
} & %
\parbox[c][1\totalheight]{3cm}{%
$[1]\to k^2 + 2p_0\cdot k$
}\\[8ex]%
\parbox[c][1\totalheight]{3cm}{%
 Region 2 %

\includegraphics[width=2.9cm]{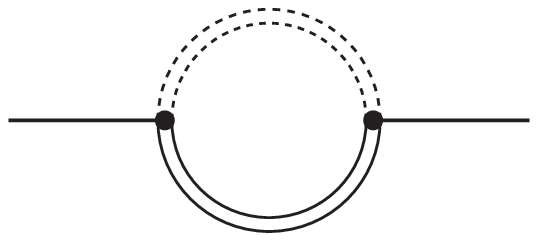}} & %
\parbox[c][1\totalheight]{2cm}{%
$|k| \sim \delta m_b$
} & %
\parbox[c][1\totalheight]{3cm}{%
$[1]\to 2p_0\cdot k + 2\delta p_0^2$
}\tabularnewline
\end{tabular}

\caption{\label{fig:asymptotic} The asymptotic expansion of the diagram in
  Fig.~\ref{fig:tree} $(a)$. The dotted double line corresponds to the epsilon
  propagator and the double solid line corresponds to the eikonal propagator
  $2p_0\cdot k + 2\delta p_0^2$}
\end{figure}

The remaining loop momenta can have one of two
characteristic scales, hard  $m_b$ or soft $m_b-m_c =\delta \cdot m_b$.
Depending on their configuration, the propagators can be expanded in
some small parameter, leading to a factorized product of one or more
single one-scale integrals
\cite{Tkachov:1997gz,Czarnecki:1996nr,Smirnov:2002pj}.  
This procedure is illustrated in
Fig.~\ref{fig:asymptotic} with the lowest-order example without gluons.

In this example, Region 1 contributes only to the real part, hence
need not be considered in this calculation of the decay rate.  
More generally, the hard regions (when all momenta are hard, $\sim m_b$) will
not contribute even at higher orders in $\alpha_s$. 
This removes what would otherwise be the most difficult part of the
calculation.
In the present expansion around $\delta=0$ there are fewer regions that need to be
considered than in the expansion around $\rho=0$.
At most four regions contribute to a given
diagram, compared to eleven in the complementary expansion of 
Refs.~\cite{Pak:2008qt,Pak:2008cp}. All other regions are either purely real
or scaleless and give no contribution. 

The second region, thus, contains full information about the tree level result. 
The resulting integral has already been considered in the
literature~\cite{Czarnecki:1997fc}.
Here we treat its generalization
since it will be  needed in the higher order corrections,
\begin{eqnarray}
  I(\lambda_1,\lambda_2,\lambda_3) &=& \int \frac{{\rm d}^Dk}{(2\pi)^D}
  \frac{1}{(k^2)^{\lambda_1} (2p_0\cdot k)^{\lambda_2} (2p_0\cdot k +
    \Delta)^{\lambda_3}}
  \label{eq:eOne} \\
  &=& (-p_0^2)^{\lambda_1-D/2} \Delta^{D-2\lambda_1-\lambda_2-\lambda_3}
  \frac{\Gamma(D-2\lambda_1-\lambda_2)
    \Gamma(2\lambda_1+\lambda_2+\lambda_3-D)
    \Gamma(D/2-\lambda_1)}{(4\pi)^{D/2} \Gamma(\lambda_1) \Gamma(\lambda_3)
    \Gamma(D-2\lambda_1)}\,,
\end{eqnarray}
where $\Delta = 2\delta p_0^2 - i0$, $p_0^2 = -m_b^2$, and
$\lambda_1$, $\lambda_2$, and $\lambda_3$ are arbitrary complex
numbers. In general, there would also be scalar products in the
numerator, but it is well known how to deal with these
\cite{Czarnecki:1997fc} and bring the integral into the form of
Eq.~(\ref{eq:eOne}). This integral is one of only five general
integrals needed  in NLO and NNLO calculations. The other
ones are on-shell propagators up to two loops and one-loop massless
propagators, all of which are well known \cite{Broadhurst:1991fi}.

For every topology, the most complicated integrals were encountered in the
regions where all loop momenta are soft. Fortunately, these three-loop
integrals could easily be written as a combination of nested integrals of the
form $I(\lambda_1,\lambda_2,\lambda_3)$. To illustrate this, let us consider
the general two-loop integral, corresponding to Fig.~\ref{fig:tree} $(b)$
after integration of the lepton loop. If both loop momenta are soft, the
integral is given by
\begin{equation}
  \int \frac{{\rm d}^Dk_1}{(2\pi)^D} \frac{{\rm d}^Dk_2}{(2\pi)^D}
  \frac{(k_1\cdot k_2)^{a_6}}{(k_1^2)^{\epsilon+a_1} (2p_0\cdot k_1 +
    \Delta)^{a_2} (k_2^2)^{a_3} (2p_0\cdot k_2)^{a_4}
    [2p_0\cdot(k_1+k_2)+\Delta]^{a_5}} \,,
   \label{eq::eTwo}
\end{equation}
where the $a_i$ are integer numbers and $a_6$ is always positive. The $k_2$
integral can be carried out using Eq.~(\ref{eq:eOne}) with $\Delta$ replaced
by $2p_0\cdot k_1 + \Delta$, performing tensor reduction in the process. The
resulting $k_1$ integral is again of the type of Eq.~(\ref{eq:eOne}).

For the NNLO calculation, it turned out to be useful to apply partial fraction
decomposition in some cases. For example, in Eq.~(\ref{eq::eTwo}) we could
also use the identity
\begin{equation}
  \frac{1}{(2p_0\cdot k_2)\, [2p_0\cdot(k_1+k_2)+\Delta]} =
  \frac{1}{(2p_0\cdot k_1 + \Delta)} \left( \frac{1}{2p_0\cdot k_2} -
    \frac{1}{2p_0\cdot(k_1+k_2)+\Delta} \right)
  \label{eq::partfrac}
\end{equation}
to reduce the number of terms in the denominator. Note that the $k_2$ integral
becomes scaleless for $a_5 \leq 0$. While it is obviously not necessary to
apply Eq.~(\ref{eq::partfrac}) in the case of the integral in
Eq.~(\ref{eq::eTwo}), it was necessary to apply analogous identities in order
to write some of the NNLO integrals as nested integrals of the type of
Eq.~(\ref{eq:eOne}).

New types of integrals appear only in the diagrams with three-gluon
interactions (cf. Fig~\ref{fig::dias} $(c)$), due to the third gluon
propagator. However, in these cases it was possible to apply the so-called
Laporta algorithm~\cite{Laporta:1996mq,Laporta:2001dd} to dispose of one of
the three gluon propagators. The remaining integrals were again a nested set
of $I$-type integrals. For this reduction we used the {\tt C++} program {\tt
  rows}~\cite{rows} and the {\tt Mathematica} package {\tt
  FIRE}~\cite{Smirnov:2008iw}.

Our calculation was performed with two independent setups. One approach is
based on the code developed for the calculation of Ref.~\cite{Pak:2008qt}. The
other uses {\tt QGRAF}~\cite{Nogueira:1991ex} to generate the diagrams and
{\tt q2e} and {\tt exp}~\cite{Harlander:1997zb,Seidensticker:1999bb} to
process them further (no expansion is done in this step). The final
calculations are in both cases done with custom code written in {\tt
  FORM}~\cite{Vermaseren:2000nd}.

\section{Results}
The result for the total width can be cast into the form
\begin{eqnarray}
  \Gamma &=& \frac{G_F^2\,|V_{cb}|^2\,m_b^5}{192\,\pi^3}\, \left[ X_0 +
  \frac{\alpha_s(m_b)}{\pi}\,C_F\,X_1 + \left( \frac{\alpha_s}{\pi}
  \right)^2\,C_F\,X_2 + \dots \right]\,,
  \label{eq::gamma}\\
  X_2 &=& C_F\,X_F + C_A\,X_A + T_F \left( n_l\,X_l + X_c + X_b \right) \,,
\end{eqnarray}
where $G_F$ is the Fermi constant and the ellipsis denotes higher order
contributions. $\alpha_s$ is defined with five active flavors. In QCD we have
$C_F = 4/3$, $C_A = 3$, and $T_F = 1/2$. $n_l = 3$ denotes the number of light
quark flavors, which are taken to be massless in our calculation. $X_c$ and
$X_b$ denote the contribution from self-energy diagrams with closed $c$- and
$b$-quark loops, respectively (cf. Fig~\ref{fig::dias} $(a)$). Thus, $X_c$
also contains contributions from real $c$-quark pairs. The quark masses are
renormalized in the on-shell scheme.

The tree level and one-loop contributions can be inferred
from the closed-form result of Ref.~\cite{Nir:1989rm}. Expanded in
$\delta$ they read
\begin{eqnarray}
  X_0 &=& \frac{64}{5}\del{5} - \frac{96}{5}\del{6} +
  \frac{288}{35}\del{7} + \dots\,, \label{eq::X0} \\
  X_1 &=& -\frac{48}{5}\del{5} + \frac{72}{5}\del{6} + \left(
  -\frac{158152}{11025} + \frac{512}{105}\lnd \right)\del{7} + \dots\,,
  \label{eq::X1}
\end{eqnarray}
where the ellipses denote higher order terms. Note that the expansion
starts at the fifth power of $\delta$. Thus, the total width tends to
zero very fast as $\delta$ ($\rho$) tends to zero (one). Logarithms of
$\delta$ always appear as $\lnd$, since they stem solely from $\Delta$ in the
integral of Eq.~(\ref{eq:eOne}).

The first three terms in the expansion of the individual contributions
read
\begin{eqnarray}
  X_F &=& \left[ -\frac{46}{5} + \frac{32}{5}\pi^2 \left( 1 - \ln2
  \right) + \frac{48}{5}\zeta_3 \right]\del{5} + \left[ \frac{69}{5} -
  \frac{48}{5}\pi^2 \left( 1 - \ln2 \right) - \frac{72}{5}\zeta_3
  \right]\del{6} \nonumber\\
  && + \left( \frac{39329}{3675} + \frac{3044}{945}\pi^2 -
  \frac{496}{105}\pi^2\ln2 + \frac{248}{35}\zeta_3 - \frac{352}{105}\lnd
  \right)\del{7} + \dots\,,
  \label{eq::cf}
\\
  X_A &=& \left[ -\frac{286}{15} - \frac{8}{5}\pi^2 \left( 1 - 2\ln2
  \right) - \frac{24}{5}\zeta_3 \right]\del{5} + \left[ \frac{99}{5} +
  \frac{12}{5}\pi^2 \left( 1 - 2\ln2 \right) + \frac{36}{5}\zeta_3
  \right]\del{6} \nonumber\\
  && + \left( -\frac{99547507}{1157625} + \frac{62206}{33075}\pi^2 +
  \frac{248}{105}\pi^2\ln2 + \frac{132}{35}\zeta_3 +
  \frac{1333376}{33075}\lnd - \frac{256}{315}\pi^2\lnd -
  \frac{1408}{315}\lndd \right)\del{7} + \dots\,, \nonumber\\
  \label{eq::ca}
\end{eqnarray}
\begin{eqnarray}
  X_l &=& \frac{56}{15}\del{5} - \frac{12}{5}\del{6} + \left[
  \frac{25577548}{1157625} - \frac{417664}{33075}\lnd +
  \frac{512}{315}\left( \lndd - \frac{\pi^2}{3} \right) \right]\del{7} +
  \dots\,,
  \label{eq::nl}
\\
  X_c &=& \left( \frac{184}{3} - \frac{32}{5}\pi^2 \right)\del{5} +
  \left( -\frac{828}{5} + \frac{88}{5}\pi^2 \right)\del{6} + \left(
  \frac{108580}{567} - \frac{18968}{945}\pi^2 \right)\del{7} + \dots\,,
  \label{eq::nc}
\\
  X_b &=& \left( \frac{184}{3} - \frac{32}{5}\pi^2 \right)\del{5} +
  \left( -12 + \frac{8}{5}\pi^2 \right)\del{6} + \left(
  \frac{107444}{2835} - \frac{3848}{945}\pi^2 \right)\del{7} + \dots\,.
  \label{eq::nb}
\end{eqnarray}
We have calculated the fermionic contributions $X_l$, $X_c$, and $X_b$
through terms of order $\del{15}$, while we computed terms of order
$\del{12}$ and $\del{11}$ for the abelian and non-abelian
contributions. Higher order terms are not shown for brevity, but are
available among the source files of this paper in arXiv.

\begin{figure}
  \includegraphics[width=0.6\textwidth]{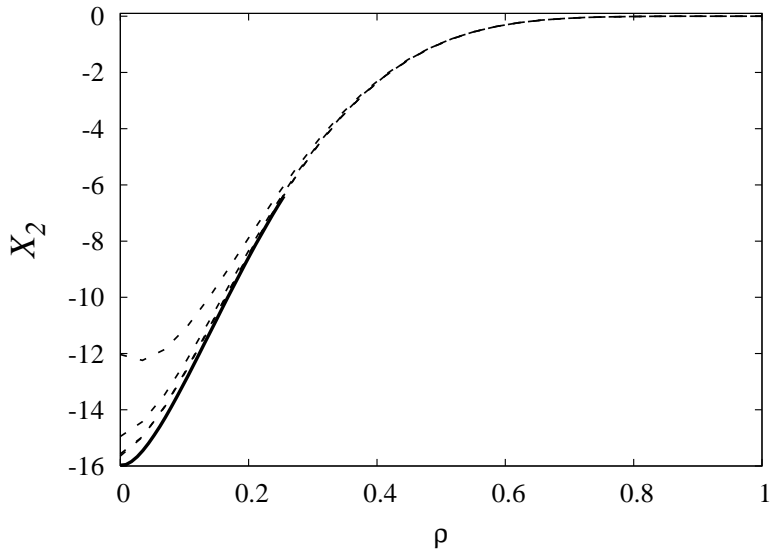}
  \caption{\label{fig::X2}$X_2$ as a function of $\rho$. The solid line
  denotes the result of Ref.~\cite{Pak:2008qt}. The dashed lines denote
  our expansion through order $\del{8}$, $\del{9}$, $\del{10}$, and
  $\del{11}$. The latter two are almost indistinguishable. (On the vertical
  axis, zero does not coincide with the upper edge of the plot.)}
\end{figure}
To illustrate the convergence behavior of our result, Fig.~\ref{fig::X2} shows
the full NNLO contribution, $X_2$, as a function of $\rho$. It shows the
expansion truncated at different orders in $\delta$ compared to the result of
Ref.~\cite{Pak:2008qt}. The latter is only given up to $\rho=0.255$, which is
were the results are closest. The convergence behavior of the expansion around
$\rho=0$ was studied in Ref.~\cite{Pak:2008cp}. Due to the suppression of our
expansion at small values of $\delta$, the different curves are
indistinguishable for $\rho > 0.4$. However, the convergence behavior is very
good even close to $\rho = 0$. This is in contrast to the expansion of
Ref.~\cite{Pak:2008qt}, which tends to $\pm\infty$ as $\rho$ tends to one.

\begin{figure}
  \includegraphics[width=0.45\textwidth]{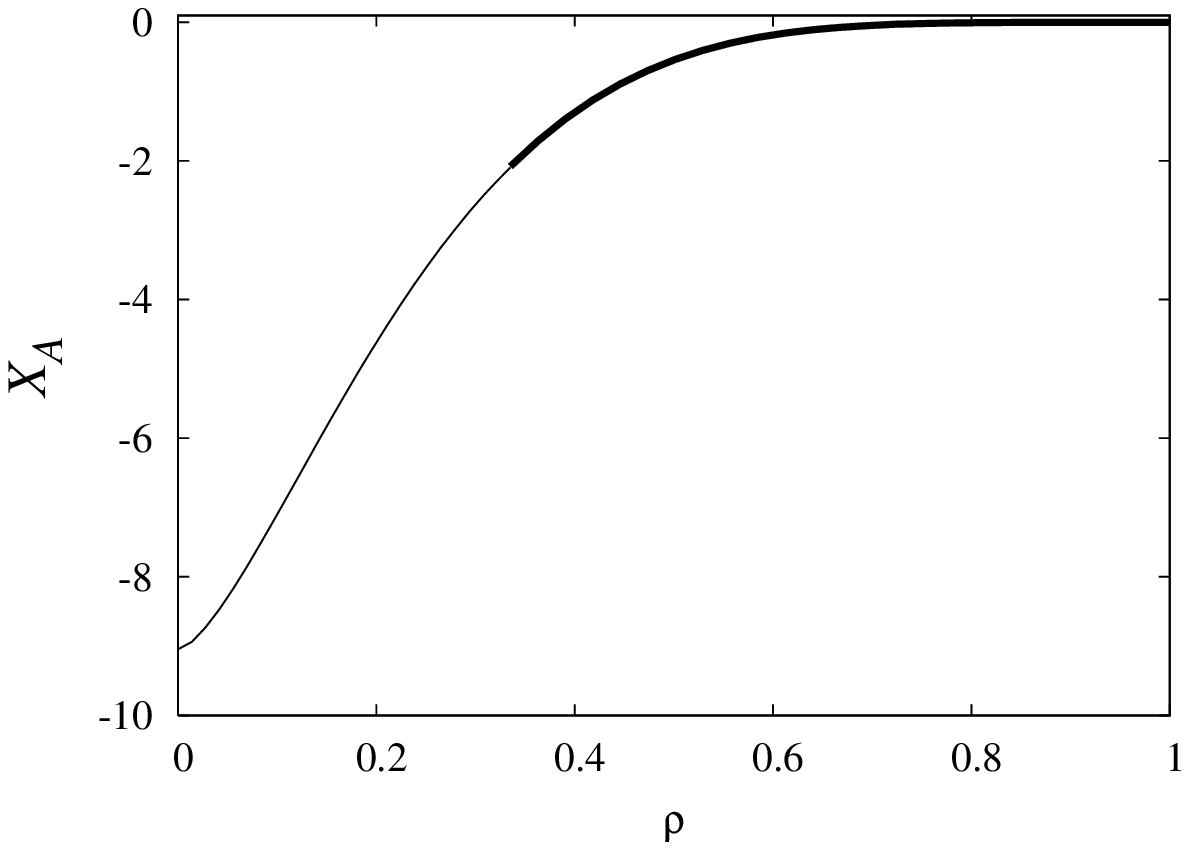}
  \includegraphics[width=0.45\textwidth]{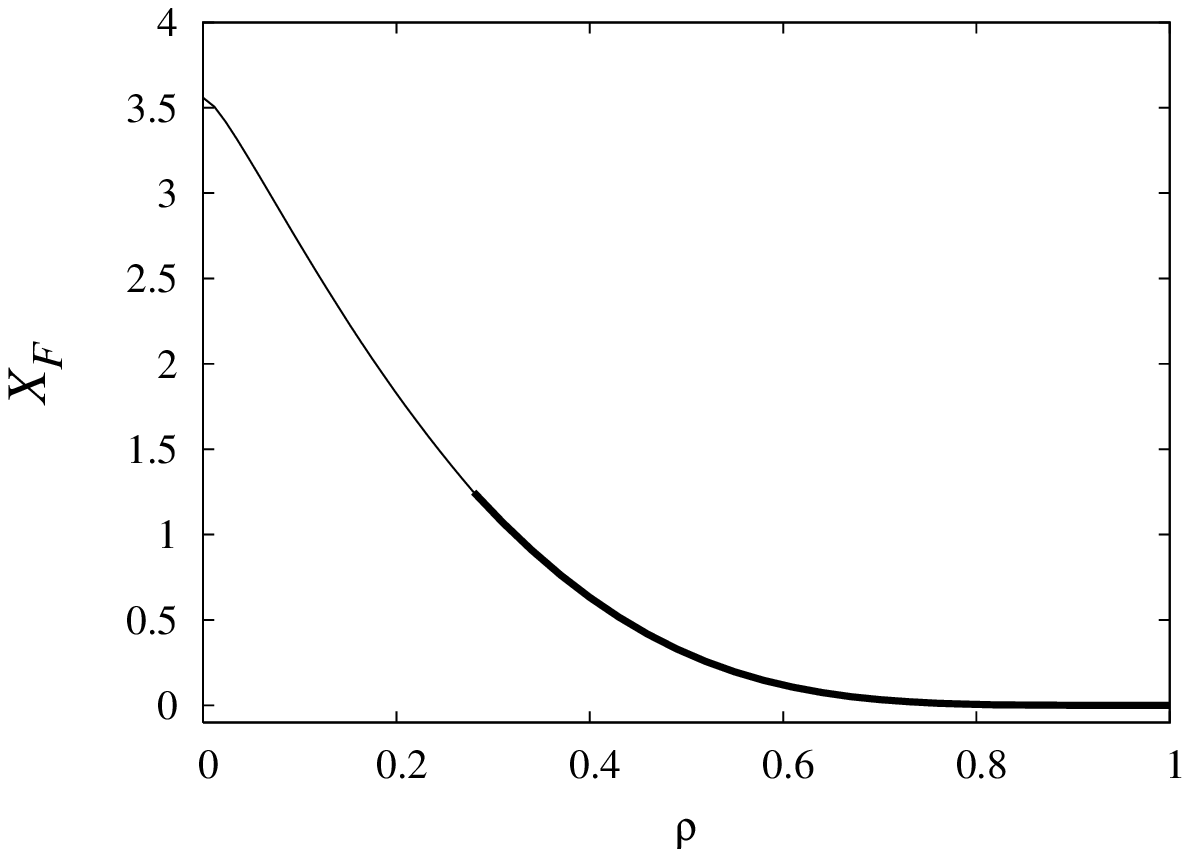}\\
  \includegraphics[width=0.45\textwidth]{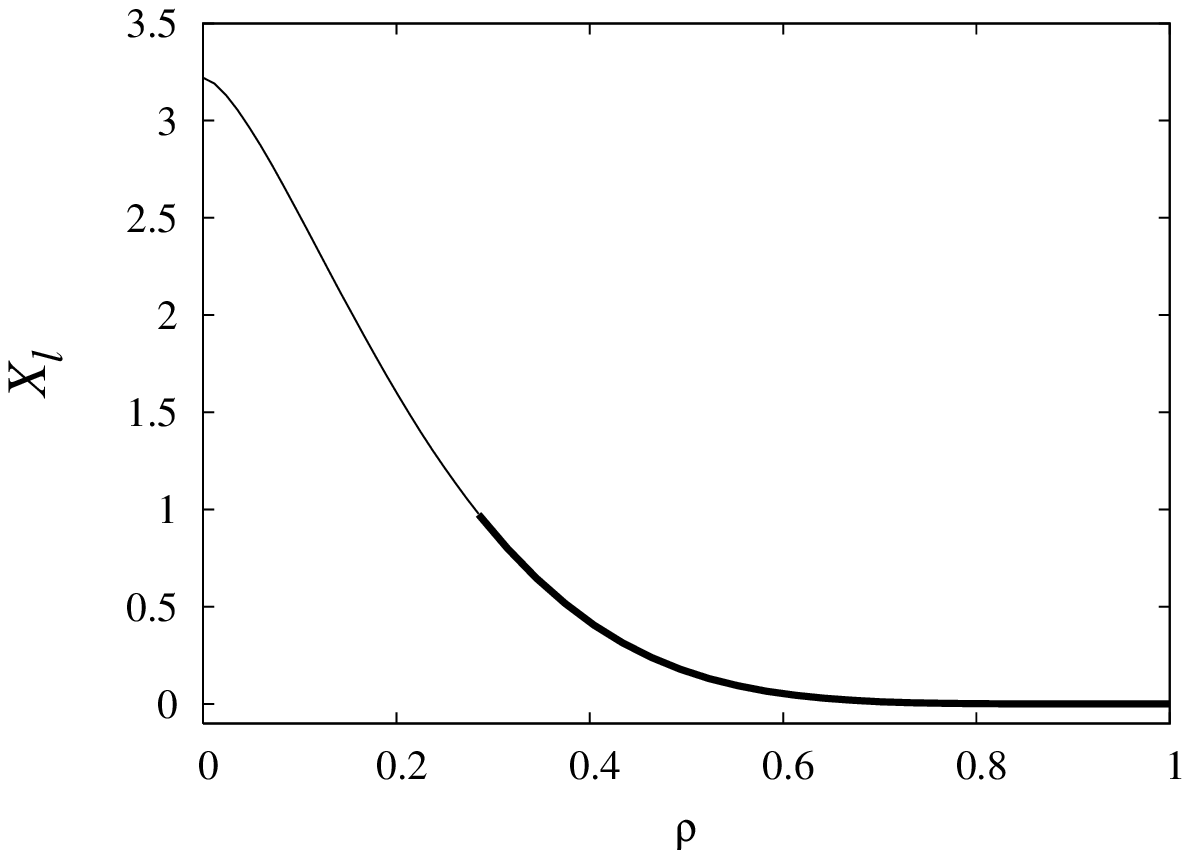}
  \includegraphics[width=0.45\textwidth]{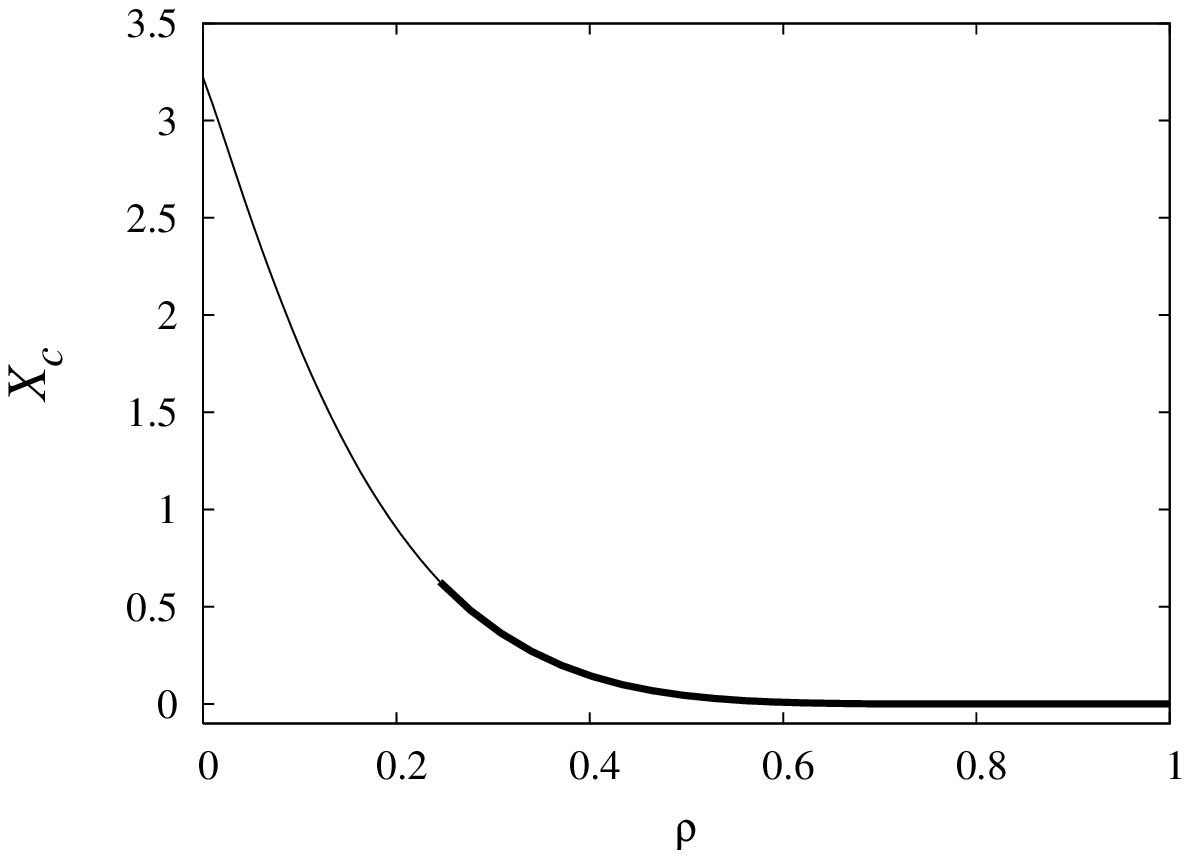}\\
  \includegraphics[width=0.45\textwidth]{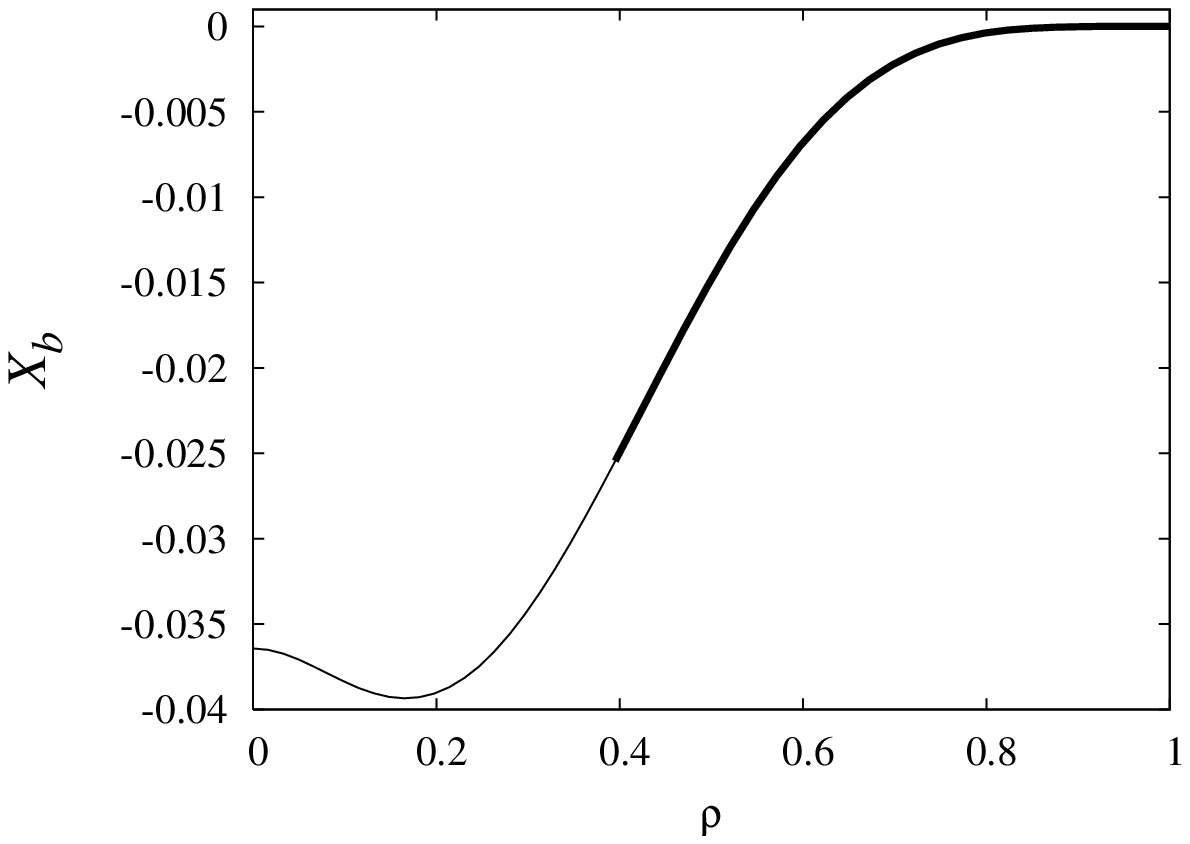}
  \caption{\label{fig::comp}NNLO contributions of the different color
  structures of the total width as functions of $\rho$. Thick and thin
  lines denote the results of Eqs.~(\ref{eq::cf})--(\ref{eq::nb}) and
  Ref.~\cite{Pak:2008qt}, respectively. Note that our expansion tends to zero
  very fast as $\rho$ tends to one. (On the vertical axis, zero does not
  coincide with the edge of the plot.)}
\end{figure}
Fig.~\ref{fig::comp} compares the different color structures with the
expansion of Refs.~\cite{Pak:2008qt,Pak:2008cp}. As the transition point between
the two results, we chose the point were they are closest. The two expansions
match very well for $\rho$ between 0.2 and 0.4. Thus, a combination of the two
results enables us to describe the decay over the whole range of kinematically
allowed values of the daughter quark mass.  It was noted in
Ref.~\cite{Pak:2008cp} that the contribution from closed $b$-quark loops shows
an extremum around $\rho = 0.2$ (cf. the last panel in
Fig.~\ref{fig::comp}). Using our expansion through $\del{15}$, we were able to
verify this behavior.

\section{Connection with the zero-recoil form factor}
In this Section, we provide an independent derivation of the first two terms in the $\delta$ expansion  of the $b\to c$
decay width.  They are independent of the real gluon radiation. The
real radiation is suppressed by the square of the velocity of the
daughter quark and influences only the third order terms, of relative
$\order{\delta^{2}}$. The first two terms, $\order{\delta^{0,1}}$,
are determined by the form factors $\eta_{A,V}$ describing the $W$-boson
coupling to quarks. Those form factors arise from virtual corrections
and replace $1-\gamma_{5}$ in that coupling by $\eta_{V}-\eta_{A}\gamma_{5}$.
The decay rate expanded in $\delta$ is, in the lowest two orders,
fully described by these form factors,
\begin{equation}
\Gamma
=\frac{G_{F}^{2}\left|V_{cb}\right|^{2}m_{b}^{5}
\left(\eta_{V}^{2}+3\eta_{A}^{2}\right)}{60\pi^3 }\, \delta^5
\left(1-\frac{3}{2}\delta + \order{\delta^{2}} \right).
\label{eq:ffRate}\end{equation}
Without strong interactions, $\eta_{V}=\eta_{A}=1$ and we reproduce
the first two terms of Eq.~(\ref{eq::X0}).

Both form factors are functions of $q^{2}$, where $q$ is the four-momentum
released in the decay. Thus, to be precise, we should have used certain
average values in Eq.~(\ref{eq:ffRate}). However, when the quark masses
are close to each other, the variation of $q^{2}$ is of the second
order in $\delta$ and can be neglected in our approximation. 

Even at a fixed $q^{2}$, the form factors depend on the difference
of the quark masses. However, in our approximation it is sufficient
to know them in the limit of equal quark masses:
because of the symmetry $m_{b}\leftrightarrow m_{c}$, the linear
term in $\delta$ vanishes, and the dependence on the quark mass difference
starts only with the quadratic term. In this limit $\eta_{V}$ equals
one to all orders, while $\eta_{A}$ is modified by the strong interactions
at $\order{\alpha_{s}}$ and higher orders. Those corrections were
calculated in Ref.~\cite{Czarnecki:1996gu} with two-loop accuracy, and
in Ref.~\cite{Archambault:2004zs} at three loops. (Full $q^{2}$ dependence
at two loops can be found in
Refs.~\cite{Bernreuther:2004ih,Bernreuther:2004th,Bernreuther:2005rw,Bernreuther:2005gq}.)

In order to compare our results with those of Ref.~\cite{Czarnecki:1996gu}, we
have to change the renormalization scale of $\alpha_s$. While we used
$\alpha_s(m_b)$ in Eq.~(\ref{eq::gamma}), Ref.~\cite{Czarnecki:1996gu} uses
$\alpha_s(\sqrt{m_cm_b})$. Note that a mistake in the running of $\alpha_s$ in
Ref.~\cite{Czarnecki:1996gu} was pointed out in Ref.~\cite{Dowling:2008ap}. In
the running from $m_b$ to $\sqrt{m_cm_b}$, five instead of four active flavors
were used. To correct for this mistake, we run $\alpha_s$ in the result of
Ref.~\cite{Czarnecki:1996gu} from $\sqrt{m_cm_b}$ to $m_b$, using five
flavors. At the scale $m_b$, we decouple the $b$ quark and run back to
$\sqrt{m_cm_b}$, using now four active flavors. Thus, the correct result is
obtained by adding
\begin{equation}
  f_{A,V}(\delta) = \frac{1}{3}\, \ln (1-\delta)\, \eta^{(1)}_{A,V} \,.
  \label{eq::fAV}
\end{equation}
to $\eta^H_{A,V}$ in Ref.~\cite{Czarnecki:1996gu}. In our comparison the
correction term contributes to the term of relative order $\order{\delta}$.

For completeness we provide all terms of Ref.~\cite{Czarnecki:1996gu} which
are needed for the comparison with our result. The QCD corrections to the
axial form factor are defined as
\begin{eqnarray}
  \eta_A &=& 1 + \frac{\alpha_s(\sqrt{m_cm_b})}{\pi}\, C_F\, \eta^{(1)}_A
  + \left( \frac{\alpha_s}{\pi} \right)^2 C_F\, \eta^{(2)}_A
  + \order{\alpha_s^3}\,, \\
  \eta^{(2)}_A &=& C_F\eta^{F}_A + \left( C_A - 2C_F \right) \eta^{AF}_A + T_F
  \left( n_l\, \eta^L_A + \eta^H_A \right) \,,
\end{eqnarray}
where $\alpha_s$ is defined with four active flavors. $\eta^H_A$ combines the
contributions from diagrams with closed $c$- and $b$-quark loops. The
individual color structures in the limit $\delta\to 0$ are given by
\begin{eqnarray}
  \eta^{(1)}_{A} & = & - \frac{1}{2} + \order{\del{2}} \,, \\
  \eta^{F}_A & = & - \frac{373}{144} + \frac{1}{6}\pi^2 + \order{\del{2}} \,,
  \\
  \eta^{AF}_{A} & = & - \frac{143}{144} - \frac{1}{12}\pi^2 +
  \frac{1}{6}\pi^2\ln 2  - \frac{1}{4}\zeta(3) + \order{\del{2}} \,, \\
  \eta_A^L &=& \frac{7}{36} + \order{\del{2}}\,, \\
  \eta^{H}_A & = & \frac{115}{18} - \frac{2}{3}\pi^2 + \frac{\delta}{6} +
  \order{\del{2}} \,.
\end{eqnarray}
The term of order $\delta$ in $\eta^H_A$ is due to the correction term
in Eq.~(\ref{eq::fAV}).  This linear term arises because the
$m_b\leftrightarrow m_c$ symmetry is broken by the charge
renormalization, since the $b$-quark does not contribute to the
running between $m_b$ and $\sqrt{m_cm_b}$.

To compare the two results, we decouple the $b$ quark in our result and run
$\alpha_s$ from $m_b$ to $\sqrt{m_cm_b}$ with four active flavors. This
changes $X_2$ by
\begin{equation}
 \delta X_2 = \frac{48}{5} \left[ \frac{11}{12} C_A - \frac{1}{3} T_F \left(n_l + 1
    \right) \right]\, \del{6} + \order{\del{7}} \,.
\end{equation}
We find perfect agreement for the first two terms of our expansion.
Comparing the widths calculated with Eqs. (\ref{eq::gamma}) and
(\ref{eq:ffRate}), we expect and indeed confirm that
\begin{equation}
\frac{\eta_A^2}{20}\left( \delta^5 - \frac{3}{2}\delta^6\right) 
=
\left.
\frac{1}{192}
\left[ \frac{3}{4}X_0
+\frac{\alpha_s(\sqrt{m_cm_b}) }{\pi} C_F X_1
+\left(\frac{\alpha_s}{\pi}\right)^2 C_F \left( X_2 + \delta X_2 \right) \right]
\right|_{\delta^{5,6}}.
\end{equation}
In the tree-level term $X_0$, the factor $3/4$ eliminates the
contribution of the vector coupling.

Individual parts in Eqs.~(\ref{eq::cf}-\ref{eq::nb}) are also reproduced;
$\eta_H^A$ combines effects of both heavy quarks and corresponds to the sum of
$X_c$ and $X_b$.

\section{Summary}
To summarize, we have calculated the semileptonic $b\to c$ decay as an
expansion around the limit of equal quark masses. Our result is a fast
convergent series, which smoothly matches the expansion in the opposite
limit. Our result confirms the calculations of
Refs.~\cite{Melnikov:2008qs,Pak:2008qt}. Together with the result of
Ref.~\cite{Pak:2008qt}, we now have analytical results valid over the whole
range of kinematically allowed daughter quark masses. An additional check of
a part of our result is afforded by comparing with the result of
Ref.~\cite{Czarnecki:1996gu}.

Furthermore, we have explained the application of the method of asymptotic
expansion to a new kinematic limit. This limit leads to significant
calculational simplifications and results in a fast convergent series, which
is applicable over most of the allowed region of the daughter quark mass. Even
in the massless limit the error is only about 2.5\% for the total width, as
demonstrated in Fig.~\ref{fig::X2}. Thus, this new expansion provides a
convenient tool for future studies of various aspects of decays not only of
quarks, but also leptons such as the muon.

\begin{acknowledgments}
We thank Alexey Pak for collaboration at an early stage of this project, 
for many helpful discussions and for sharing with us his program {\tt rows}. 
This work is supported by  Science and Engineering Research
Canada. The Feynman diagrams were drawn with {\tt
  JaxoDraw}~\cite{Binosi:2003yf}. 
\end{acknowledgments}

\end{document}